\documentclass[aps,prb,twocolumn,superscriptaddress,raggedbottom,showpacs,floatfix]{revtex4-1}
\usepackage{amsmath,amsfonts,amssymb,amsthm}
\usepackage{color}
\usepackage{graphicx,latexsym}

\begin{document}
\renewcommand{\vec}{\mathbf}
\renewcommand{\Re}{\mathop{\mathrm{Re}}\nolimits}
\renewcommand{\Im}{\mathop{\mathrm{Im}}\nolimits}

\title{Fluctuational internal Josephson effect in topological insulator film}
\author{D.K. Efimkin}
\affiliation{Institute of Spectroscopy  RAS, 142190, Troitsk, Moscow, Russia}
\author{Yu.E. Lozovik}
\altaffiliation{email: lozovik@isan.troitsk.ru}

\affiliation{Institute of Spectroscopy RAS, 142190, Troitsk, Moscow, Russia}
\affiliation{Moscow Institute of Physics and Technology, 141700,
Moscow, Russia}

\begin{abstract}
Tunneling between opposite surfaces of topological insulator thin film populated by electrons and holes is considered. We predict considerable enhancement of tunneling conductivity by Cooper electron-hole pair fluctuations that are precursor of their Cooper pairing. Cooper pair fluctuations lead to the critical behavior of tunneling conductivity in vicinity of critical temperature with critical index $\nu=2$. If the pairing is suppressed by disorder the behavior of tunneling conductivity in vicinity of quantum phase transition is also critical with the index $\mu=2$. The effect can be interpreted as fluctuational internal Josephson effect and it is general phenomenon for electron-hole bilayers. The peculiarities of the effect in other realizations of electron-hole bilayer are discussed.
\end{abstract}
\maketitle

\section{Introduction}
Cooper pairing of spatially separated electrons and holes was predicted in the system of semiconductor quantum wells more then thirty years ago\cite{LozovikYudson}.  Later it was observed in quantum Hall bilayer at total filling factor $\nu_{\mathrm{T}}=1$ that can be presented as the system of spatially separated composite electrons and composite holes (see \cite{EisensteinMacDonald} and references therein). After graphene discovery Cooper pairing of Dirac electrons and holes in the structure of independently gated graphene layers has been proposed \cite{LozovikSokolik,MinBistrizerSuMacDonald,KharitonovEfetov}. Recently possibility of Cooper pairing of Dirac electrons and holes was predicted\cite{EfimkinLozovikSokolik,SeradjehMooreFranz} in thin film of topological insulator (TI), new unique class of solids that has topologically protected Dirac surface states \cite{HasanKane,QiZhang}. The electron-hole pairing in that system is the realization of topological superfluidity and hosts Majorana fermions on edges and vortices \cite{SeradjehMooreFranz,Seradjeh} that is the topic of extraordinary interest due to possibility to use them in quantum computation \cite{Alicea,Beenakker}. Also the Cooper pairing can lead to a number of interesting physical effects including superfluidity \cite{LozovikYudson}, anomalous drag effect\cite{VignaleMacDonald}, nonlocal Andreev reflection \cite{PesinMacDonald}.

The most prominent manifestation of electron-hole pairing is internal Josephson effect \cite{LozovikPoushnov}. The coherence between electron and hole states leads to a tunnel current $j_{\mathrm{T}}=j_0 \sin(\phi-\phi_{\mathrm{T}})$ that depends on the phase $\phi$ of electron-hole condensate. Here $j_0$ is maximal value of the current carried by the condensate and $\phi_{\mathrm{T}}$ is the phase of the tunnel matrix element. Dynamic of the phase on the macroscopic scale is described by the action with Lagrangian analogous to the one for the superconducting Josephson junctions. But current-voltage characteristics of the electron-hole bilayer drastically differs from ones of the latter. In equilibrium state the phase of the order parameter is fixated $\phi=\phi_{\mathrm{T}}$ and the tunnel current is zero hence in electron-hole system there is no analog of DC Josephson effect. The coherent tunnel current flows in the non equilibrium state driven by a voltage bias between the electron and hole layers. It leads to a colossal enhancement of tunneling conductivity at zero voltage bias. The effect has been observed in quantum Hall bilayer\cite{JosehsonExp1,JosehsonExp2} and its microscopical and macroscopical description were addressed in a number of interesting theoretical papers \cite{SternGirvinMacDonaldMa,JoglekarMacDonald,RosLee,FoglerWilczek,BezuglyjShevchenko}.

Cooper electron-hole pairing can appear above critical temperature as thermodynamic fluctuations.  Particulary they lead to the logarithmic divergence of a drag conductivity as a function of a temperature\cite{Hu,Mink} and a pseudogap formation in single-particle density of states of electrons and holes\cite{Rist}. Manifestations of Cooper pair fluctuations in tunneling have not been consider previously. Since tunneling conductivity is colossally enhanced in the paired state one can anticipate its strong enhancement by Cooper pair fluctuations above critical temperature in analogy with strong contribution of electron-electron Copper pair fluctuations in superconductor to its diamagnetic susceptibility and electric conductivity \cite{LarkinVarlamov}. Indeed, tunneling current in electron-hole bilayer can be transferred by  Cooper pair fluctuations. Since the amplitude of pairing fluctuations increases in a vicinity of critical temperature, as fluctuations of an ordered state do for different phase transitions, one can expect the significant enhancement of tunneling conductivity and its critical behavior. The described effect can be called fluctuational internal Josephson effect and it is rather general phenomenon for electron-hole bilayers. Here we develop the microscopic theory of the effect and its macroscopic theory will be published elsewhere. We have considered the effect in topological insulator thin film and its peculiarities in other realizations of electron-hole bilayer are discussed in Conclusions.

The rest of the paper is organized as follows. In Section 2 we briefly discuss the model used for the description of interacting electrons and holes in TI film. In Section 3 the microscopical description of Cooper pair fluctuations is introduced. In Section 4 the tunneling conductivity between the opposite surfaces of topological insulator film is calculated and Section 5 is devoted to the analysis of results and conclusions.\begin{figure}[t]
\label{Fig1}
\begin{center}
\includegraphics[width=8.5 cm]{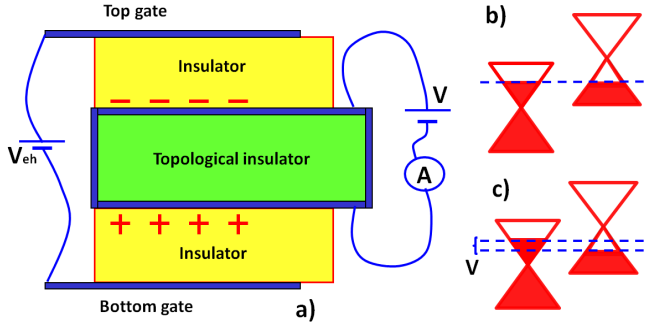}
\caption{(Color online) a) Experimental setup for measurements of tunneling conductivity between opposite surfaces of TI populated by electrons and holes. b) Dispersion laws of the electrons and holes in equilibrium. Dashed line denotes  electrochemical potential of TI surfaces. c) Dispersion law of the electrons and holes in nonequilibrium state induced by voltage bias V. }
\end{center}
\end{figure}
\section{The model}
The  setup for the experimental investigation of tunneling conductivity between spatially separated electrons and holes in TI film is presented on Fig.1. Voltage $V_{\mathrm{eh}}$ between the external gates induces equilibrium concentrations of electrons and holes on the opposite surfaces. Voltage $V$ drives the system from equilibrium and induces charge current between the layers. If the side surfaces of the film are gapped, for example, by ordered magnetic impurities introduced to TI surface the charge can be transferred only via interlayer tunneling and the tunneling resistance can be measured. Also charge transport through TI side surfaces is unimportant if the area of the tunneling junction is large enough.

Possibility and peculiarities of electron-hole Cooper pairing in the TI film in realistic model that takes into account screening, disorder and interlayer tunneling has been considered in our paper \cite{EfimkinLozovikSokolik}.  Here we focus on investigation of Cooper pair fluctuations and their role in tunneling. Hamiltonian of the system $H=H_{\mathrm{eh}}+H_{\mathrm{d}}+H_{\mathrm{T}}$ includes kinetic and electron-hole interaction energies $H_\mathrm{eh}$, interaction with disorder $H_{\mathrm{d}}$ and the part describing tunneling $H_{\mathrm{T}}$. The first part in single-band approximation that ignores valence (conduction) band on the surface with excess of electrons (holes) is given by
\begin{equation}
\begin{split}
H_{\mathrm{eh}}=&\sum_{\vec{p}}\xi_{\vec{p}} a_\vec{p}^+a_\vec{p} - \sum_{\vec{p}}\xi_{\vec{p}} b_\vec{p}^+b_\vec{p} + \\ &+\sum_{\vec{p}\vec{p}'\vec{q}} U(\vec{q})\Lambda_{\vec{p}'-\vec{q},\vec{p}'}^{\vec{p}+\vec{q},\vec{p}}a_{\vec{p}+\vec{q}}^+b^+_{\vec{p}\prime-\vec{q}}
b_{\vec{p}\prime}^{\mathstrut} a_{\vec{p}}^{\mathstrut}.
\end{split}
\end{equation}
Here $a_{\vec{p}}$ is annihilation operator for a electron on the surface with excess of electrons and  $b_{\vec{p}}$ is annihilation operator for a electron on the surface with excess of holes \cite{Comment1}; $\xi_{\vec{p}}=v_{\mathrm{F}}p-E_{\mathrm{F}}$ is Dirac dispersion law in which $v_{\mathrm{F}}$ and $E_{\mathrm{F}}$ are velocity and Fermi energy of electrons and holes. We consider the balanced case since it is favorable for Cooper pairing and the pairing is sensitive to concentration mismatch of electrons and holes. $U(\vec{q})$ is screened Coulomb interaction between electrons and holes (see \cite{EfimkinLozovikSokolik} for its explicit value) and $\Lambda_{\vec{p}'-\vec{q},\vec{p}'}^{\vec{p}+\vec{q},\vec{p}} =
\cos{\small (\phi_{\vec{p},\vec{p}+\vec{q}}/2)} \cos{(\phi_{\vec{p}',\vec{p}'+\vec{q}}/2)}$ is angle factor that comes from the overlap of spinor wave functions of two-dimensional Dirac fermions. Critical temperature of pairing in Bardeen-Cooper-Schrieffer (BCS) theory that ignores disorder and tunneling is given by
\begin{equation}
T_0=\frac{2\gamma'E_{\mathrm{F}}}{\pi} \exp^{-1/\nu_\mathrm{F}U'}
\end{equation}
where $U'$ is Coulomb coupling constant \cite{Comment2} ; $\gamma'=e^C$ where $C\approx0,577$ is the Euler constant.
\begin{figure}[t]
\label{Fig2}
\begin{center}
\includegraphics[width=8.7 cm]{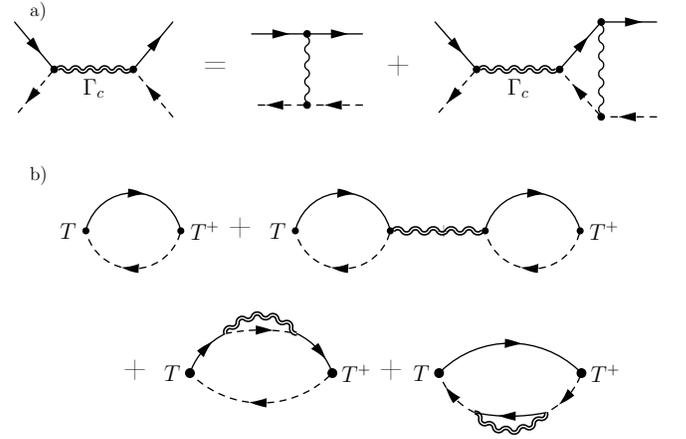}
\caption{a) Diagrammatic representation of the Bethe-Salpeter equation for the Cooper propogator $\Gamma_\mathrm{c}$; b) Feynman diagrams for a tunnel conductivity. Solid (dashed) line corresponds to electrons on the surface of TI film with excess of electrons (holes).}
\end{center}
\end{figure}

We do not specify explicitly the interaction Hamiltonian with disorder $H_\mathrm{d}$ since both short-range and long-range Coulomb impurities lead to pairbreaking and can suppress Cooper pairing. Short-range disorder scatters only one component of Cooper pair since they are spatially separated and long-range Coulomb impurities acts differently on components of Cooper pair since they have different charge. Below we introduce phenomenological decays of electrons $\gamma_\mathrm{a}$ and holes $\gamma_\mathrm{b}$.

The tunneling of electrons between the opposite surfaces of topological insulator thin film with conserving momentum can be described by the following Hamiltonian
\begin{equation}
H_{\mathrm{T}}=T+T^+=\sum_{\vec{p}} \left(t b_{\vec{p}}^+a_{\vec{p}}+t^* a_{\vec{p}}^+b_{\vec{p}}\right),
\end{equation}
where $t$ is the tunneling amplitude. We consider influence of tunneling on pairing to be weak and treat it below as perturbation.

The described model is applicable for description of tunneling in TI films which width is larger than value at which $t\approx T_0$. If $t \gg T_0$ tunneling strongly influences electron-hole pairing and it can not be treated as perturbation. Particularly it induces electron-hole condensate with fixated phase and smears critical temperature to the paired state. Our calculation \cite{EfimkinLozovikSokolik} shows that the described model is applicable for thin films of $\hbox{Bi}_2 \hbox{Se}_3$ at $d> 10\;\hbox{nm}$. In that case critical temperature without disorder can achieve $T_0\approx0.1\;\hbox{K}$ at $E_{\mathrm{F}}=5 \;\hbox{meV}$ and the pairing is not suppress by disorder if electrons and holes have exceptional hight mobilities of order $\mu\sim 10^6 \;\hbox{sm}^2/\hbox{V}\hbox{s}$. It should be noted that single-band and static screening approximations used for calculation of critical temperature\cite{EfimkinLozovikSokolik} usually underestimate critical temperature of Cooper pairing between Dirac particles \cite{LozovikOgarkovSokolik,LozovikSokolikMultiband,SodemannPesinMacDonald}.

\section{Cooper pair fluctuations}
For the microscopical description of Cooper pair fluctuations we introduce Cooper propagator $\Gamma_\mathrm{c}^{\mathrm{R}}(\omega)$. It corresponds to the two-particle vertex function in the Cooper channel \cite{LarkinVarlamov} and satisfies the Bethe-Salpeter equation depicted on Fig. 2 (a). In Bardeen-Cooper-Schrieffer (BCS) approximation its solution can be presented in the form
\begin{equation}\label{CooperPropagator}
\Gamma_\mathrm{c}^{\mathrm{R}}(\omega)=\frac{U'}{1-U'\Pi_{\mathrm{c}}^{\mathrm{R}}(\omega)},
\end{equation}
where $\Pi_{\mathrm{c}}^{\mathrm{R}}(\omega)$ corresponds to electron-hole bubble diagram that can be interpreted as Cooper susceptibility of the system. After direct calculation it can be presented in the following form
\begin{equation}
\label{Bubble}
\begin{split}
\Pi_\mathrm{c}^\mathrm{R}(\omega)=\frac{1}{U'}-\nu_\mathrm{F}\ln\frac{T}{T_0} - \nu_\mathrm{F} \Psi\left(\frac{1}{2}\right) - \\ - \nu_\mathrm{F}\Psi\left(\frac{1}{2}-\frac{i\omega}{4 \pi T}+\frac{\gamma}{2\pi T}\right).
\end{split}
\end{equation}
Here $\gamma=(\gamma_\mathrm{a}+\gamma_\mathrm{b})/2$ is disorder caused Copper pair decay rate equals to the half-sum of phenomenological introduced decay rates of electrons and holes\cite{Comment3};  $\nu_{\mathrm{F}}$ is the density of states of electrons and holes on the Fermi level; $\Psi(x)$ is the digamma function. Cooper pair propagator acquires the following form
\begin{equation}
\label{CooperPropagator2}
\Gamma_\mathrm{c}^{\mathrm{R}}(\omega)=\frac{1}{\nu_{\mathrm{F}}}\frac{1}{\ln\frac{T}{T_\mathrm{0}}+\Psi\left(\frac{1}{2}-i\frac{\omega}{4 \pi T}+\frac{\gamma}{2\pi T}\right)-\Psi\left(\frac{1}{2}\right)}.
\end{equation}
In the absence of disorder $\Gamma_\mathrm{c}^{\mathrm{R}}(\omega)=0$ at the critical temperature $T_0$ indicating Cooper instability of the system against Cooper pairing. Critical temperature for disordered system $T_\mathrm{d}$ at which $\Gamma_\mathrm{c}^{\mathrm{R}}(\omega)=0$ satisfies the following equation
\begin{equation}
\ln\frac{T_\mathrm{d}}{T_\mathrm{0}}+\Psi\left(\frac{1}{2}+\frac{\gamma}{2\pi T}\right)-\Psi \left(\frac{1}{2}\right)=0.
\end{equation}
This equation has nontrivial solution if $\gamma<\gamma_0$, where $\gamma_{0}=0.89 T_0$ is the critical Cooper pair decay value. In opposite case the pairing is suppressed by disorder. The value $\gamma_0$ corresponds to quantum critical point at zero temperature.

Above critical temperature $T_{\mathrm{d}}$ the expression for Cooper pair propagator (\ref{CooperPropagator2}) at $\omega\rightarrow 0$ can be approximated in the following way
\begin{equation}
\Gamma_\mathrm{c}^{\mathrm{R}}(\omega)=\frac{1}{\nu_{\mathrm{F}}}\frac{4\pi T_{\mathrm{d}}}{4\pi T_{\mathrm{d}}\ln \frac{T}{T_{\mathrm{d}}}-i\omega\Psi^\prime(\frac{1}{2}+\frac{\gamma}{2\pi T_{\mathrm{d}}})}.
\end{equation}
If the pairing is suppressed by the disorder Cooper pair propagator at zero temperature and at  $\omega\rightarrow 0$ is given by
\begin{equation}
\Gamma_\mathrm{c}^{\mathrm{R}}(\omega)=\frac{1}{\nu_{\mathrm{F}}}\frac{2 \gamma}{2\gamma\ln\frac{\gamma}{\gamma_0} - i \omega}.
\end{equation}
\begin{figure}[t]
\label{Fig3}
\begin{center}
\includegraphics[width=8.5 cm]{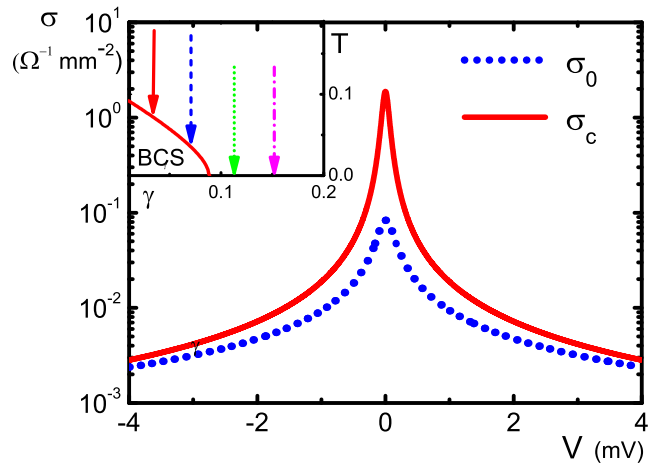}
\caption{(Color online) Tunneling conductivity $\sigma$ as a function of bias voltage $V$ for noninteracting (dashed line) and interacting (solid line) electrons and holes for $\gamma=0.2\;\hbox{K}$ and $T=0.2\;\hbox{K}$. Inset: The phase diagram of the system in which BCS denotes the pairing state. The arrows on the phase diagram correspond to the dependencies on Fig. 4-7.}
\end{center}
\end{figure}
\section{Tunneling conductivity}
For a calculation of the tunneling conductivity we use linear response theory in which the tunneling conductivity $\sigma(V)$ at a finite voltage bias $V$ can be presented in the form of Kubo formula\cite{Mahan}
\begin{equation}
\sigma(V)=\frac{e^2}{h} \frac{4 \pi}{eV} \Im[\chi^\mathrm{R}(eV)],
\end{equation}
where the retarded response function $\chi^\mathrm{R}(\omega)$ can be obtained by the analytical continuation $i\Omega_n\rightarrow \omega+i\delta$ of $\chi^\mathrm{M}(i \Omega_n)$ that is given by
\begin{equation}
\label{TunnelDef}
\chi^\mathrm{M}(i \Omega_n)=-\frac{1}{2\beta}\int_{-\beta}^{\beta}d\tau\;e^{i\Omega_n\tau}\langle T_{\mathrm{M}}T(\tau)T^+(0)\rangle.
\end{equation}
Here $T_{\mathrm{M}}$ is the time-ordering symbol for a imaginary time $\tau$ and $\Omega_n=2\pi n T$ is a bosonic Matsubara frequency. In the system of noninteracting electrons and holes the $\chi^\mathrm{M}(i \Omega_n)$ corresponds to the first diagram on the Fig. 1 (b) leading to $\chi^\mathrm{R}(\omega)=|t|^2\Pi_{\mathrm{c}}^{\mathrm{R}}(\omega)$. Hence the tunneling conductivity for noninteracting electrons and holes $\sigma_{0}$ is given by
\begin{equation}
\label{TunnelBare}
\sigma_{0}(V)=\frac{e^2}{h} \frac{4 \pi |t|^2}{eV} \Im[\Pi_{\mathrm{c}}^\mathrm{R}(eV)].
\end{equation}
Its value $\sigma_{0}^{\mathrm{m}}(\gamma,T)$ at zero bias is given by
\begin{equation}
\label{SigmaBare}
\sigma_{0}^{\mathrm{m}}(\gamma,T)=\frac{2 \pi e^2}{h} \frac{\nu_{\mathrm{F}}|t|^2}{2\pi T} \Psi'\left(\frac{1}{2}+\frac{\gamma}{2\pi T}\right),
\end{equation}
and at low temperatures $T\ll\gamma$ it transforms to
\begin{equation}
\label{igmaBareZero}
\sigma_{0}^{\mathrm{m}}(\gamma,0)=\frac{2 \pi e^2}{h} \frac{\nu_{\mathrm{F}}|t|^2}{\gamma}.
\end{equation}
\begin{figure}[t]
\label{Fig4}
\begin{center}
\includegraphics[width=8.7 cm]{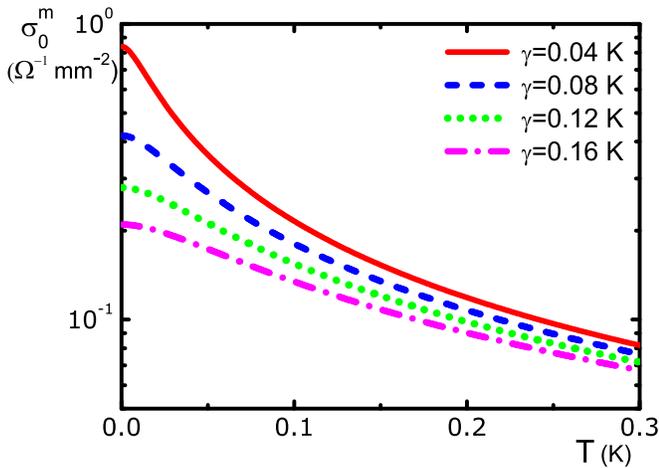}
\caption{(Color online)  The height of the tunnel conductivity peak $\sigma_{0}^{\mathrm{m}}$ for noninteracting  electrons and holes on temperature $T$ for different values of Cooper pair decay $\gamma$.}
\end{center}
\end{figure}
Introduction of the electron-hole Coulomb interaction in the ladder approximation leads to three additional terms for the tunneling conductivity. The first one corresponds to second diagram on Fig.2 (b). It is singular in a vicinity of critical temperature and cannot be reduced to the tunneling conductivity of noninteracting quasiparticles with a renormalized spectrum due to Cooper pair fluctuations. The other two terms correspond to renormalization of single-particle Green functions of electrons and holes. They are not singular in a vicinity of critical temperature and can be neglected. The tunneling conductivity for interacting electrons and holes $\sigma_{\mathrm{c}}$ is given by
\begin{equation}
\label{TunnelInt}
\sigma_{\mathrm{c}}(V)=\frac{e^2}{h} \frac{4 \pi |t|^2}{eV} \Im\left[\frac{\Pi_\mathrm{c}^\mathrm{R}(eV)}{1-U'\Pi_{\mathrm{c}}^\mathrm{R}(eV)}\right].
\end{equation}
The denominator of (\ref{TunnelInt}) coincides with that in the Cooper propagator (\ref{CooperPropagator}) and tends to zero in a vicinity of critical temperature $T_{\mathrm{d}}$. Hence the Cooper pair fluctuations lead to the critical behavior of tunneling conductivity in vicinity of the critical temperature and the quantum critical point. Above critical temperature $T_{\mathrm{d}}$  the tunneling conductivity at zero bias is given by
\begin{equation}
\label{SigmaFluctClassical}
\sigma_\mathrm{c}^{\mathrm{m}}(\gamma,T)=\sigma_\mathrm{0}^{\mathrm{m}}(\gamma,T)\frac{1}{(\nu_\mathrm{F}U^\prime)^2}\frac{1}{\ln^2 \frac{T}{T_\mathrm{d}}}.
\end{equation}
In vicinity of the critical temperature it diverges as $\sigma_\mathrm{c}^{\mathrm{m}}(\gamma,T)\sim(T-T_\mathrm{d})^{-\nu}$ with the critical index $\nu=2$. At zero temperature tunneling conductivity is given by
\begin{equation}
\label{SigmaFluctQuantum}
\sigma_{\mathrm{c}}^{\mathrm{m}}(\gamma,0)=\sigma_{0}^\mathrm{m}(\gamma,0)\frac{1}{(\nu_{\mathrm{F}}U')^2} \frac{1}{\ln^2\frac{\gamma}{\gamma_0}}.
\end{equation}
It diverges in the vicinity of the quantum phase transition at $\gamma=\gamma_0$ as $\sigma_\mathrm{c}^{\mathrm{m}}(\gamma,0) \sim(\gamma -\gamma_0)^{-\mu}$ with the critical index $\mu=2$.

The formulas (\ref{SigmaFluctClassical}) and (\ref{SigmaFluctQuantum}) are the main result of the article. The calculated contribution of Cooper pair fluctuations to the tunneling conductivity cannot be reduced to the one of noninteracting quasiparticles with renormalized spectrum due to Cooper pair fluctuations. It can be interpreted as the \emph{direct} contribution of Cooper pair fluctuations. Hence the predicted effect of the enhancement of tunneling conductivity in a vicinity of the critical temperature and the quantum critical point is the collective effect that can be interpreted as fluctuational internal Josephson effect.
\begin{figure}[t]
\label{Fig5}
\begin{center}
\includegraphics[width=8.7 cm]{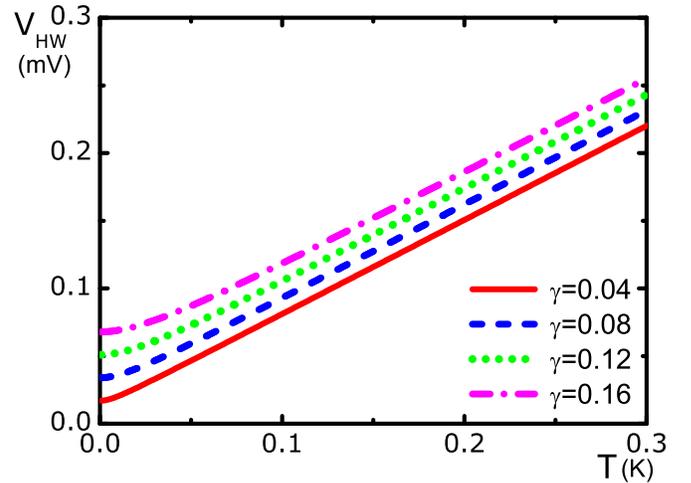}
\caption{(Color online) Half-width of the tunnel conductivity peak $V_\mathrm{HW}$ for noninteracting electrons and holes as function of temperature $T$ for different values of Cooper pair decay $\gamma$.}
\end{center}
\end{figure}
\section{Analysis and Discussion}
Tunneling conductivity at finite voltage bias and in full-range of temperature and Cooper pair decay was calculated numerically according to formulas (\ref{Bubble}), (\ref{TunnelBare}) and (\ref{TunnelInt}). The following set\cite{EfimkinLozovikSokolik} of the parameters $T_0=0.1\;\hbox{K}$, $E_{\mathrm{F}}=5\;\hbox{meV}$, $\nu_{\mathrm{}F}U'=0.16$, $t=10\;\mu\hbox{eV}$ was used. The set corresponds to $\hbox{Bi}_2 \hbox{Se}_3$ TI film with width $10 \;\hbox{nm}$. The phase diagram of the system is presented on the inset of Fig. 3.  If Cooper pair decay rate exceeds critical value $\gamma_0=0.09\;\hbox{K}$ then the electron-hole pairing is suppressed by a disorder.

Calculated tunneling conductivity both for noninteracting electrons and holes and interacting ones for $\gamma=0.2 \; \hbox{K}$ and $T=0.2 \; \hbox{K}$ is presented on Fig. 3. The dependence has prominent peak and is qualitatively the same for all points of the phase diagram $(\gamma,T)$. The peak appears due to restrictions connected with energy and momentum conservation for tunneling electrons. Such peak was predicted and observed also in electron-electron bilayers \cite{Tunneling2D2DExp,ZhengMacDonald} and it is the peculiarity of a tunneling between two two-dimensional systems. Coulomb interaction between electrons and holes considerably enhances the tunneling conductivity but does not change qualitatively its dependence on external bias.
\begin{figure}[t]
\label{Fig6}
\begin{center}
\includegraphics[width=8.7 cm]{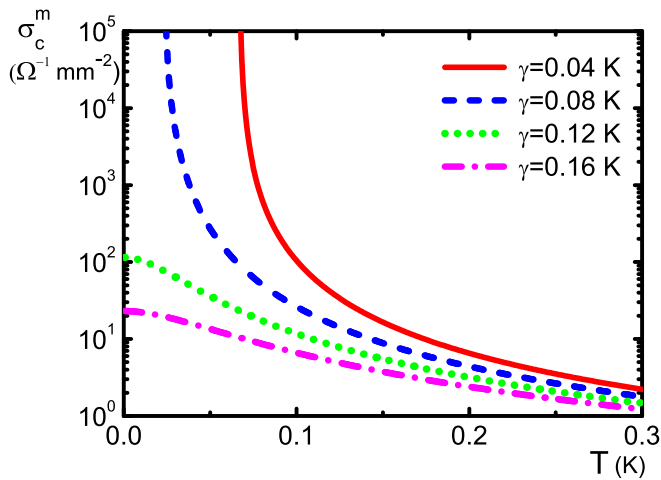}
\caption{(Color online)  The height of the tunnel conductivity peak $\sigma_{\mathrm{c}}^{\mathrm{m}}$ for interacting  electrons and holes on temperature $T$ for different values of Cooper pair decay $\gamma$. }
\end{center}
\end{figure}

Height and half-width of the peak for noninteracting electrons are presented on Figs. 4 and 5. The peak becomes more prominent with decreasing of Cooper pair decay and temperature. The peaks width is determined by $\max\{\gamma,T\}$. The peaks height is determined as $1/\max\{\gamma,T\}$ and it is decreasing as $1/T$ at $T\gg\gamma$. The height and half-width of the peak for interacting electrons and holes are presented on Fig. 6 and Fig. 7, respectively. For $\gamma<\gamma_0$ Coulomb interaction leads to the critical behavior of the tunneling conductivity in a vicinity of the critical temperature that we interpret as fluctuational internal Josephson effect. In vicinity of the critical temperature height of the peak diverges with the critical index $\nu=2$ that agrees with the analytic results and the width of the peak linearly tends to zero. At high temperatures peaks height decreases as $1/(T\log^2 (T/T_\mathrm{d}))$. The critical region in which tunneling conductivity is considerably enhanced is of order $\Delta T\approx T_d$.  If the Cooper pair decay exceeds the critical value $\gamma>\gamma_0$ Coulomb interaction considerably enhances the height and leads to reduction of the width but does not lead to any singularities. The peak becomes more prominent with decreasing of decay and temperature as it does in the model of noninteracting electrons and holes. The peaks height smoothly depends on temperature but its maximal value at zero temperature diverges as function of Cooper pair decay $\gamma$ in vicinity of quantum critical point at $\gamma_0$. The width of the critical region is of order $\Delta \gamma \approx \gamma_0$.

The peaks width and height smoothly depend on the parameters of the system used for the calculation and listed above. But a satisfaction of the number of assumptions is important for observation of fluctuational internal Josephson effect .

The model we use here is well applicable in the regime of weak hybridization $t \ll T_0$  in which influence of tunneling on Cooper pairing can be neglected. In ultrathin TI film the regime of strong hybridization $t\gg T_0$  can be realized. In that regime tunneling induces the gap $t$ in the spectrum of electrons and holes which is considerable larger than the one due to their Cooper pairing. Critical temperature and Cooper instability are considerably smoothed in that case. So in that regime we do not expect critical behavior of the tunneling conductivity due to Cooper pair fluctuations. Our calculations for $\hbox{Bi}_2 \hbox{Se}_3$ shows that regime of strong hybridization can be realized in films which width is less then $10\; \hbox{nm}$.

\begin{figure}[t]
\label{Fig7}
\begin{center}
\includegraphics[width=8.9 cm]{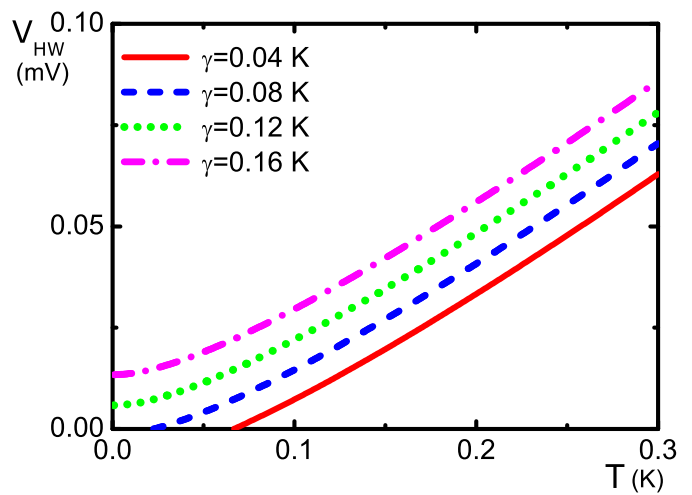}
\caption{(Color online)  Half-width of the tunnel conductivity peak $V_\mathrm{HW}$ for interacting electrons and holes as function of temperature $T$ for different values of Cooper pair decay rate $\gamma$.}
\end{center}
\end{figure}

The mean field theory we use here for the description of fluctuational internal Josephson effect does not account large scale fluctuation of phase of Cooper pair condensate. In two-dimensional superfluids phase fluctuations destroy long-range coherence and the transition to paired state at critical temperature $T_{\mathrm{d}}$ calculated within mean field theory is smoothed. Moreover the transition to superfluid state is Berezinskii-Kosterlitz-Thouless transition \cite{Berezinskii,KosterlizThouless} that corresponds to dissociation of vortex-antivortex pairs and which temperature is below $T_{\mathrm{d}}$. Hence the large scale phase fluctuations of Cooper pair condensate can smooth the critical behavior of tunnel conductivity we predict here. But if the size of the system is comparable with coherence length of Cooper pair fluctuations  $l_{\mathrm{c}}\approx\hbar v_{\mathrm{F}}/T_0$ the phase fluctuations are unimportant and mean field theory is well applicable. For $T_0=0.1 \; \hbox{K}$ the coherence length of Cooper pair fluctuations is of order $l_{\mathrm{c}}\sim 10 \; \hbox{mkm}$ and we conclude that the developed microscopical theory is applicable for samples of the corresponding size.

The model we use here implies conservation of the momentum of tunneling electron. If the momentum is not conserved the tunneling process creates electron-hole pair with nonzero total momentum of order $l_{\mathrm{T}}^{-1}$. Here $l_{\mathrm{T}}$ is character length at which tunneling matrix matrix element $t$ can be considered as constant. Cooper pair is formed by electron and hole with opposite momenta and the Cooper instability is smoothed if $l_{\mathrm{c}}\gg l_{\mathrm{T}}$. For tunneling between opposite surfaces of topological insulator thin film of high crystalline quality momentum conservation can be achieved to remarkable degree.

We have shown that electron-hole Coulomb interaction considerably enhances the tunneling conductivity in electron-hole bilayer even when the Cooper pairing is suppressed by disorder. The opposite situation takes place in electron-electron bilayer that also can be realized in semiconductor quantum well structure, in graphene double layer system and in a film of topological insulator. Coulomb interaction gives contribution to decay of electrons that was analyzed in \cite{JungwirthMacDonald} and to additional series of diagrams for the tunneling conductivity. We treated the additional diagrams in the ladder approximation (See Fig.2-b). If they are omitted the tunneling conductivity at zero temperature and at a finite bias $V$ is given by \cite{JungwirthMacDonald,ZhengMacDonald}
\begin{equation}
\sigma_{0}^{\mathrm{ee}}=gt_{\mathrm{A}}\frac{2\pi e^2}{h}\frac{\nu_{\mathrm{F}}|t|^2}{\gamma} \frac{4\gamma^2}{4\gamma^2+(eV)^2},
\end{equation}
where $g$ is the degeneracy factor of electrons and $t_{\mathrm{A}}$ is additional factor that depends on internal nature of electrons\cite{Comment4}. The dependence of tunnel conductivity on external bias contains prominent peak which becomes more prominent with decreasing of decay rate $\gamma$. If the Coulomb interaction is treated in ladder approximation the tunneling conductivity is given by
\begin{equation}
\sigma_{\mathrm{c}}^{\mathrm{ee}}=gt_{\mathrm{A}}\frac{2\pi e^2}{h}\frac{\nu_{\mathrm{F}}|t|^2}{\gamma} \frac{4\gamma^2}{4\gamma^2+(1+(\nu_{\mathrm{F}}U')^2)(eV)^2}.
\end{equation}
For electron-electron bilayer Coulomb interaction does not influences the height of the peak and leads to decreasing of the width which is insignificant even in the case of strong interaction $\nu_{\mathrm{F}}U'\sim1$. The roles of the interlayer Coulomb interaction in electron-electron bilayer and electron-hole bilayer are drastically different because the correction to the tunneling conductivity of the primer is caused by the scattering diagrams in particle-antiparticle channel and the correction to the one of the latter is caused by the diagrams in particle-particle Cooper channel that contains instability.

We have investigated the manifestations of Cooper electron-hole pairing fluctuations in thin film of topological insulator on tunneling between its opposite surfaces. The internal fluctuational Josephson effect is general phenomenon but each realization of electron-hole bilayer has its own peculiarities.

Dirac points in graphene are situated in corners of first Brillouin zone. Electron-hole pairing was predicted\cite{LozovikSokolik,MinBistrizerSuMacDonald,KharitonovEfetov} in the system of two independently gated graphene layers separated by dielectric film. In that case orientations of the graphene lattices are uncorrelated. The distance between Dirac points of different layers in momentum space is of order $a_0^{-1}$, where $a_0$ is lattice constant of graphene. The tunneling of electrons between Dirac points is possible if $l_{\mathrm{T}}\sim a_0$ that corresponds to tunneling through impurity states or other defects. The condition $l_{\mathrm{c}}\gg l_{\mathrm{T}}$ is well satisfied and the critical behavior of tunneling conductivity in double layer graphene system is considerably smoothed. But if the mutual orientation of graphene layers can be controlled in experiment the presented here theory is well applicable for that system. The formulas (\ref{SigmaFluctClassical}),(\ref{SigmaFluctQuantum}) are reasonable and can be easily generalized to additional spin and valley degree of freedom of electrons and holes. So the fluctuational internal Josephson effect can also be experimentally investigates in that system.

Recently anomalies in drag effect in semiconductor double well structure that contains spatially separated electrons and holes were observed \cite{CroxallExp,MorathExp}. The analysis of results shows that the observed anomalies can be caused by electron-hole pairing not in BCS regime but rather in regime of BCS-BEC crossover\cite{Leggett,PieriNeilsonStrinati}. In that regime electron-hole pairing fluctuations also should increase tunneling conductivity of the system but quantitative theory of the effect is interesting and
challenging problem. The developed here microscopical theory of fluctuation internal Josephson effect is applicable for that system if the pairing is realized in BCS regime. Moreover the formulas (\ref{SigmaFluctClassical}),(\ref{SigmaFluctQuantum}) are reasonable in that case. In the semiconductor double well structures\cite{CroxallExp,MorathExp} the concentrations of electrons and holes can be independently controlled and separate contacts to the layers have been made. So in that system tunneling conductivity between electron and hole layers can be measured and the predictions of our work can be also addressed.

Internal Josephson effect has been observed in quantum Hall bilayer at total occupation factor $\nu_{\mathrm{T}}=1$. But above critical temperature the dependence of tunneling conductivity on external bias does not contain any peak due to non Fermi-liquid behavior of composite electrons and holes \cite{QHFTunnelingExp1,QHFTunnelingExp2,QHFTunnelingTheor1,QHFTunnelingTheor2}. Thus fluctuational effects in that system are more complicated ones and need separate investigation. It should be noted that critical behavior of tunneling conductivity has not been observed yet in experiments in that system.

We have considered influence of electron-hole Cooper pair fluctuations that are precursor of their Cooper pairing in topological insulator film on tunneling between its opposite surface. Cooper pair fluctuations lead to critical behavior of tunneling conductivity in vicinity of critical temperature with critical index $\nu=2$. If pairing is suppressed by disorder the behavior of tunneling conductivity in vicinity of quantum critical point at zero temperature is also critical with critical index $\mu=2$. The effect can be interpreted as fluctuational Josephson effect.
\begin{acknowledgments}
The authors are indebted to A.A. Sokolik for discussions. The work was supported by RFBR programs including grant 12-02-31199. D.K.E acknowledge support from Dynasty Foundation.
\end{acknowledgments}

\bibliographystyle{apsrev}

\begin{thebibliography}{99}
\bibitem{Comment1}
We use here anihilation operators for electrons from both surfaces of topological insulator film. But for interpretation of results it is more convenient to use the language of electrons and holes.
\bibitem{Comment2}
The couling constant is determined by multipole momenta of screened Coulomb interaction on Fermi level and is given by $U'=(U_0(k_\mathrm{F},k_\mathrm{F})+U_1(k_\mathrm{F},k_\mathrm{F}))/2$. For detailes see \cite{EfimkinLozovikSokolik}.
\bibitem{Comment3}
Taking into account correlation between disorder's potentials acting on electrons and holes leads only to redifinition of paramater $\gamma$ that we treat phenomenologically here.
\bibitem{Comment4}
For conventional electrons with quadratic dispersion law in semiconductor quantum wells the factor $t_{\mathrm{A}}=1$. For Dirac electrons in thin film of topological insulator it is equal to $t_{\mathrm{A}}=0$ due to conservation of spin in tunneling process. And for Dirac electrons in graphene the factor ranges from $0$ to $1$ and depends on mutual orientations of graphene layers.
\end{thebibliography}

\end{document}